\documentclass[conference]{IEEEtran}
\IEEEoverridecommandlockouts
\usepackage{cite}
\usepackage{amsmath}
\usepackage{amsfonts}
\usepackage{amssymb}
\usepackage{url}
\usepackage{algorithm}
\usepackage{algorithmic}
\usepackage{graphicx}
\usepackage{textcomp}
\usepackage{tikz}
\usepackage{adjustbox}
\usepackage{tabularx}
\usepackage{subcaption}
\usepackage{multirow}
\usepackage{multicol}
\usetikzlibrary{quantikz2}
\usepackage[hidelinks]{hyperref}
\usepackage[capitalise]{cleveref}
\usepackage{booktabs}
\usepackage{blochsphere}
\usepackage{dsfont}
\usepackage{tablefootnote}
\creflabelformat{equation}{#2#1#3}

\def\BibTeX{{\rm B\kern-.05em{\sc i\kern-.025em b}\kern-.08em
T\kern-.1667em\lower.7ex\hbox{E}\kern-.125emX}}

\makeatletter
\newcommand{\linebreakand}{%
\end{@IEEEauthorhalign}
\hfill\mbox{}\par
\mbox{}\hfill
\begin{@IEEEauthorhalign}
}
\makeatother

\begin{document}

\title{Emergent Cooperation in Quantum Multi-Agent Reinforcement Learning Using Communication
\thanks{This work is part of the Munich Quantum Valley, which is supported by the Bavarian state government with funds from the Hightech Agenda Bayern Plus. Sponsored in part by the Bavarian Ministry of Economic Affairs, Regional Development and Energy as part of the 6GQT project.}}

\author{
  \IEEEauthorblockN{Michael Kölle} 
  \IEEEauthorblockA{\textit{LMU Munich} \\
    Munich, Germany \\
  michael.koelle@ifi.lmu.de}
  \and
  \IEEEauthorblockN{Christian Reff}
  \IEEEauthorblockA{\textit{LMU Munich} \\
    Munich, Germany \\
  christian.reff@campus.lmu.de}
  \and
  \IEEEauthorblockN{Leo Sünkel}
  \IEEEauthorblockA{\textit{LMU Munich} \\
    Munich, Germany \\
  leo.suenkel@ifi.lmu.de}
  \linebreakand
  \IEEEauthorblockN{Julian Hager}
  \IEEEauthorblockA{\textit{LMU Munich} \\
    Munich, Germany \\
  julian.hager@ifi.lmu.de}
  \and
  \IEEEauthorblockN{Gerhard Stenzel}
  \IEEEauthorblockA{\textit{LMU Munich} \\
    Munich, Germany \\
  gerhard.stenzel@ifi.lmu.de}
  \and
  \IEEEauthorblockN{Claudia Linnhoff-Popien}
  \IEEEauthorblockA{\textit{LMU Munich} \\
    Munich, Germany \\
  linnhoff@ifi.lmu.de}
}

\maketitle

\begin{abstract}
  Emergent cooperation in classical Multi-Agent Reinforcement Learning has gained significant attention, particularly in the context of Sequential Social Dilemmas (SSDs). While classical reinforcement learning approaches have demonstrated capability for emergent cooperation, research on extending these methods to Quantum Multi-Agent Reinforcement Learning remains limited, particularly through communication. In this paper, we apply communication approaches to quantum Q-Learning agents: the Mutual Acknowledgment Token Exchange (MATE) protocol, its extension Mutually Endorsed Distributed Incentive Acknowledgment Token Exchange (MEDIATE), the peer rewarding mechanism Gifting, and Reinforced Inter-Agent Learning (RIAL). We evaluate these approaches in three SSDs: the Iterated Prisoner's Dilemma, Iterated Stag Hunt, and Iterated Game of Chicken. Our experimental results show that approaches using MATE with temporal-difference measure (MATE\textsubscript{TD}), AutoMATE, MEDIATE-I, and MEDIATE-S achieved high cooperation levels across all dilemmas, demonstrating that communication is a viable mechanism for fostering emergent cooperation in Quantum Multi-Agent Reinforcement Learning.
\end{abstract}


\section{Introduction}
\label{sec:introduction}

Many real-world artificial intelligence applications in fields such as autonomous driving \cite{shalevshwartz2016safemultiagentreinforcementlearning}, robotics \cite{10.1177/0278364913495721}, and smart manufacturing \cite{KIM2020440} increasingly rely on Multi-Agent Systems. Within these systems, multiple agents interact in a shared environment and can significantly influence each other’s outcomes. Multi-Agent Reinforcement Learning (MARL) has proven effective for modeling such interactions by allowing agents to learn optimal or near-optimal strategies through direct experience. However, the non-stationary nature of multi-agent environments—where each agent’s policy updates alter the underlying dynamics—poses substantial challenges \cite{DBLP:journals/corr/abs-1906-04737}. These challenges become more pronounced when self-interested agents compete over shared resources, often foregoing cooperative actions that could yield higher collective returns in the long run.

A growing body of literature addresses the phenomenon of emergent cooperation, which involves agents shifting from purely self-interested behavior to strategies that also benefit their peers. Researchers have proposed mechanisms such as indirect reciprocity \cite{10.5555/3635637.3663196}, reputation systems \cite{orzan2023emergent}, status-quo loss functions \cite{DBLP:journals/corr/abs-2001-05458}, and memory-enhanced learning \cite{DING2023114032}. These mechanisms are often evaluated through SSDs, which capture the tension between individual and collective interests \cite{doi:10.1073/pnas.092080099}. Communication, in particular, has consistently emerged as a key enabler of cooperation in both human \cite{doi:10.1177/0022002709352443} and artificial agents \cite{10.5555/3237383.3237408,Phan_2024,icaart25,10.5555/3398761.3398855,NEURIPS2020_ad7ed5d4}.

Parallel to these developments, Quantum Reinforcement Learning (QRL) leverages quantum properties such as superposition and entanglement to improve computational efficiency and learning capacity \cite{biamonte2017quantum, Lockwood_Si_2020}. Variational Quantum Circuits (VQCs) have been applied to Q-Learning \cite{9144562, Skolik2022quantumagentsingym} and policy-gradient methods \cite{NEURIPS2021_eec96a7f, Sequeira_2023}, demonstrating competitive performance with significantly fewer parameters. In multi-agent contexts, quantum extensions have shown promise in coordination and resource efficiency \cite{K_lle_2024,10627769,derieux2025eqmarlentangledquantummultiagent}, though most work to date has focused on performance rather than cooperation.

Despite these advances, emergent cooperation in Quantum Multi-Agent Reinforcement Learning (QMARL) remains largely unexplored—particularly through explicit communication. Understanding how communication protocols affect cooperative dynamics among quantum agents is crucial for both theoretical and practical progress in QMARL.

This work bridges that gap by adapting and empirically evaluating eight established communication mechanisms—originally developed for classical MARL—for quantum Q-Learning agents. We assess their impact on cooperation within three canonical SSDs: the Iterated Prisoner’s Dilemma, Iterated Stag Hunt, and Iterated Game of Chicken. Our findings show that communication-based approaches, especially those employing MATE\textsubscript{TD}, AutoMATE, and MEDIATE, reliably foster cooperation among quantum agents, highlighting communication as a viable path toward emergent coordination in QMARL.

This paper is organized as follows: \cref{sec:communication_protocols} presents the communication methods adapted for quantum agents. \cref{sec:vqc_architecture} details the Variational Quantum Circuit architectures used in our experiments. \cref{sec:experimental_setup} outlines the experimental setup, including the environments and evaluation metrics. \cref{sec:results} reports and analyzes the results, and \cref{sec:conclusion} concludes with key findings and recommendations for future research.

\section{Communication Mechanisms}
\label{sec:communication_protocols}


\subsection{Mutual Acknowledgment Token Exchange}

MATE~\cite{Phan_2024} is a two-phase communication protocol comprising \emph{request} and \emph{response} phases. Each agent $i$ quantifies its situation quality via a \emph{monotonic improvement measure} $MI_i$, which checks whether the agent's condition is improving or deteriorating. We employ two variants \emph{MATE\textsubscript{TD}}: \(MI_i^{\text{rew}}(\hat{r}_{t,i}) = \hat{r}_{t,i} - \overline{r}_{t,i}\) and  \emph{MATE\textsubscript{rew}}: \(MI_i^{\text{TD}}(\hat{r}_{t,i}) = \hat{r}_{t,i} + \gamma [(1-\epsilon)\,\max_{a_{t+1,i}} Q_{\theta_i}(s_{t+1,i},a_{t+1,i}) + \epsilon \sum_{a_{t+1,i}} Q_{\theta_i}(s_{t+1,i},a_{t+1,i})] - [(1-\epsilon)\,\max_{a_{t,i}} Q_{\theta_i}(s_{t,i},a_{t,i}) + \epsilon \sum_{a_{t,i}} Q_{\theta_i}(s_{t,i},a_{t,i})]\).
%
%
Here, $\hat{r}_{t,i}$ is the agent's (possibly shaped) reward at time $t$, and $\overline{r}_{t,i}$ is the agent's average reward in the current episode. Since Q-Learning does not directly learn a state-value function, we approximate $V_{\pi_i}(s_{t,i})$ by taking maximum or expected Q-values in an $\epsilon$-greedy manner.

During the \emph{request phase}, each agent $i$ computes $MI_i(r_{t,i})$. If $MI_i(r_{t,i}) \ge 0$, it sends a \emph{request token} $x_j^{\text{MATE}}$ to each neighbor $j \in \mathcal{N}_{t,i}$. In the \emph{response phase}, every neighbor $j$ receiving $x_j^{\text{MATE}}$ checks whether adding the token still yields a monotonic improvement, i.e.\ $MI_j\bigl(r_{t,j} + x_j^{\text{MATE}}\bigr) \ge 0$. If so, $j$ responds with a positive token $y_i^{\text{MATE}}$; otherwise, it responds with $-y_i^{\text{MATE}}$.
After these phases, each agent $i$ shapes its reward as
$\hat{r}_{t,i}^{\text{MATE}} = r_{t,i} + \hat{r}_{t,i}^{\text{req}} + \hat{r}_{t,i}^{\text{res}}$,
where $\hat{r}_{t,i}^{\text{req}} = 0$ if no request tokens are received and otherwise
$\hat{r}_{t,i}^{\text{req}} = \max_{x \in \{x_j^{\text{MATE}}\}_{j \in \mathcal{N}_{t,i}}} x$.
Similarly, $\hat{r}_{t,i}^{\text{res}} = 0$ if no response tokens are received and otherwise
$\hat{r}_{t,i}^{\text{res}} = \min_{y \in \{y_j^{\text{MATE}}\}_{j \in \mathcal{N}_{t,i}}} y$.
%
%
%
%
In both the request and response phases, the token values $x^{\text{MATE}}$ and $y^{\text{MATE}}$ are fixed and identical for all agents.

\subsection{MEDIATE Extensions}

MEDIATE~\cite{icaart25} builds upon MATE by introducing automatic token derivation and a decentralized consensus to keep token values identical across agents. Each agent $i$ initializes its local token $\mathcal{T}_i$ to 0.1. At the end of each epoch, $\mathcal{T}_i$ is updated through \(\triangledown_{\mathcal{T}_i} = \alpha_{\text{MEDIATE}}\,
\frac{\bigl|\tilde{V}_i - \mathrm{median}(\overline{V}_i)\bigr|}{\tilde{V}_i}\,\bigl|r_i^{\min}\bigr|\),
%
%
where $\alpha_{\text{MEDIATE}}$ is a learning rate, $\tilde{V}_i$ is the median state-value estimate from the previous epoch, $\mathrm{median}(\overline{V}_i)$ is the median from the current epoch, and $r_i^{\min}$ is the smallest non-zero reward observed. After the update, $\mathcal{T}_i$ is clamped to non-negative values to ensure niceness. For consensus, each agent $i$ splits $\mathcal{T}_i$ into $|\mathcal{N}_{t,i}| + 1$ additive shares. These shares are exchanged so that each agent can reconstruct a consensus token $\mathcal{T}_i^*$. MATE's request and response phases then use this consensus token.

Altmann et al.~\cite{icaart25} differentiate three variants. The first, \emph{AutoMATE}, relies purely on the decentralized token derivation and does not perform consensus. The second, \emph{MEDIATE-I}, maintains local tokens separately but uses the consensus token $\mathcal{T}_i^*$ for MATE requests and responses. The third, \emph{MEDIATE-S}, synchronizes the local token to the consensus token before every update, thus ensuring the same token value is used by all agents. All MEDIATE variants employ the temporal-difference version $MI_i^{\text{TD}}$ for monotonic improvement checks.

\subsection{Gifting}

Originally introduced by Lupu et al.~\cite{10.5555/3398761.3398855}, \emph{Gifting} expands the action space so that one agent can directly reward another. We focus on two variants. In \emph{Gifting\textsubscript{Zerosum}}, the agent can gift an amount $x^{\text{Gift}}$ any number of times per episode, but must pay an equivalent penalty $-x^{\text{Gift}}$. In \emph{Gifting\textsubscript{Budget}}, each agent has a fixed per-episode budget $B$ which decreases by $x^{\text{Gift}}$ every time a gift is given. Once the budget is exhausted, further gifts have no effect until the next episode. Because most of our environments require an environment action every step, we implement gifting as a separate binary decision for each agent (\emph{gift} or \emph{no gift}). The agent maintains Q-values for these two gifting actions, chosen by $\epsilon$-greedy selection. In Gifting\textsubscript{Budget}, selecting \emph{gift} decreases the agent's available budget, while Gifting\textsubscript{Zerosum} imposes a penalty to the agent equal to the gift amount.

\subsection{Reinforced Inter-Agent Learning}

RIAL~\cite{10.5555/3157096.3157336} provides a discrete communication protocol learned through independent Q-Learning. Each agent $i$ sends $k$-bit messages to others, denoted $m_{t,i,j} \in \{0,1\}^k$. In the simplest case of $k=2$, there are four possible messages $(00, 01, 10, 11)$. We incorporate this by learning separate Q-values for each bit combination. The agent selects each communication bit via $\epsilon$-greedy exploration in parallel with its standard environment action. Over time, agents adapt these learned messages to enhance cooperation, especially in partially observable tasks or when explicit information exchange is critical.

\section{Variational Quantum Circuit Architecture}
\label{sec:vqc_architecture}

\begin{figure*}[htbp]
  \centering
  \subfloat[VQC with Basis embedding of observation features $o_k$ and 4 layers used for MATE\textsubscript{rew}, MATE\textsubscript{TD}, AutoMATE, MEDIATE-I and MEDIATE-S.\label{fig:Standard VQC}]{
    \begin{adjustbox}{width=0.31\linewidth}
      \begin{quantikz}
        \lstick{$\ket{0}$} & \gate{R_X(o_1)}
        \gategroup[2,steps=6,style={inner xsep=5pt, inner ysep=6pt},background]{Repeated 4 times}
        \gategroup[2,steps=1,style={dashed,rounded corners,fill=green!20,inner xsep=2pt},background,label style={label position=below,anchor=north,yshift=-0.2cm}]{Embedding}
        & \gate{R_Z(\theta_{1,1,1})}
        \gategroup[2,steps=5,style={dashed,rounded corners,fill=blue!20,inner xsep=2pt},background,label style={label position=below,anchor=north,yshift=-0.2cm}]{Variational Layer}
        & \gate{R_Y(\theta_{1,1,2})} & \gate{R_Z(\theta_{1,1,3})} & \ctrl{1} & \targ{} & \ \ldots\ & \meter{} \\
        \lstick{$\ket{0}$} & \gate{R_X(o_2)} & \gate{R_Z(\theta_{1,2,1})} & \gate{R_Y(\theta_{1,2,2})} & \gate{R_Z(\theta_{1,2,3})} & \targ{} & \ctrl{-1} & \ \ldots\ & \meter{}
      \end{quantikz}
    \end{adjustbox}
  }
  \hfill
  \subfloat[VQC for Gifting, using Basis embedding and additional Angle embedding for budget.\label{fig:Gifting Budget VQC}]{
    \begin{adjustbox}{width=0.31\linewidth}
      \begin{quantikz}
        \lstick{$\ket{0}$} & \gate{R_X(o_1)}
        \gategroup[4,steps=8,style={inner xsep=5pt,inner ysep=6pt},background]{Repeated 4 times}
        \gategroup[3,steps=1,style={dashed,rounded corners,fill=green!20,inner xsep=2pt},background,label style={label position=below,anchor=north,yshift=-1.4cm}]{Embedding}
        & \gate{R_Z(\theta_{1,1,1})}
        \gategroup[4,steps=7,style={dashed,rounded corners,fill=blue!20,inner xsep=2pt},background,label style={label position=below,anchor=north,yshift=-0.2cm}]{Variational Layer}
        & \gate{R_Y(\theta_{1,1,2})} & \gate{R_Z(\theta_{1,1,3})} & \ctrl{1} & & & \targ{} & \ \ldots\ & \meter{} \\
        \lstick{$\ket{0}$} & \gate{R_X(o_2)} & \gate{R_Z(\theta_{1,2,1})} & \gate{R_Y(\theta_{1,2,2})} & \gate{R_Z(\theta_{1,2,3})} & \targ{} & \ctrl{1} & & & \ \ldots\ & \meter{} \\
        \lstick{$\ket{0}$} & \gate{R_X(\text{Budget})} & \gate{R_Z(\theta_{1,3,1})} & \gate{R_Y(\theta_{1,3,2})} & \gate{R_Z(\theta_{1,3,3})} & & \targ{} & \ctrl{1} & & \ \ldots\ & \meter{} \\
        \lstick{$\ket{0}$} &  & \gate{R_Z(\theta_{1,4,1})} & \gate{R_Y(\theta_{1,4,2})} & \gate{R_Z(\theta_{1,4,3})} & & & \targ{} & \ctrl{-3} & \ \ldots\ & \meter{}
      \end{quantikz}
    \end{adjustbox}
  }
  \hfill
  \subfloat[VQC for RIAL with 4 layers, adding two-bit message space.\label{fig:RIAL VQC}]{
    \begin{adjustbox}{width=0.31\linewidth}
      \begin{quantikz}
        \lstick{$\ket{0}$} & \gate{R_X(o_1)}
        \gategroup[6,steps=10,style={inner xsep=5pt,inner ysep=6pt},background]{Repeated 4 times}
        \gategroup[6,steps=1,style={dashed,rounded corners,fill=green!20,inner xsep=2pt},background,label style={label position=below,anchor=north,yshift=-0.2cm}]{Embedding}
        & \gate{R_Z(\theta_{1,1,1})}
        \gategroup[6,steps=9,style={dashed,rounded corners,fill=blue!20,inner xsep=2pt},background,label style={label position=below,anchor=north,yshift=-0.2cm}]{Variational Layer}
        & \gate{R_Y(\theta_{1,1,2})} & \gate{R_Z(\theta_{1,1,3})} & \ctrl{1} & & & & & \targ{} & \ \ldots\ & \meter{} \\
        \lstick{$\ket{0}$} & \gate{R_X(o_2)} & \gate{R_Z(\theta_{1,2,1})} & \gate{R_Y(\theta_{1,2,2})} & \gate{R_Z(\theta_{1,2,3})} & \targ{} & \ctrl{1} & & & & & \ \ldots\ & \meter{} \\
        \lstick{$\ket{0}$} & \gate{R_X(o_3)} & \gate{R_Z(\theta_{1,3,1})} & \gate{R_Y(\theta_{1,3,2})} & \gate{R_Z(\theta_{1,3,3})} & & \targ{} & \ctrl{1} & & & & \ \ldots\ & \meter{} \\
        \lstick{$\ket{0}$} & \gate{R_X(o_4)} & \gate{R_Z(\theta_{1,4,1})} & \gate{R_Y(\theta_{1,4,2})} & \gate{R_Z(\theta_{1,4,3})} & & & \targ{} & \ctrl{1} & & & \ \ldots\ & \meter{} \\
        \lstick{$\ket{0}$} & \gate{R_X(o_5)} & \gate{R_Z(\theta_{1,5,1})} & \gate{R_Y(\theta_{1,5,2})} & \gate{R_Z(\theta_{1,5,3})} & & & & \targ{} & \ctrl{1} & & \ \ldots\ & \meter{} \\
        \lstick{$\ket{0}$} & \gate{R_X(o_6)} & \gate{R_Z(\theta_{1,6,1})} & \gate{R_Y(\theta_{1,6,2})} & \gate{R_Z(\theta_{1,6,3})} & & & & & \targ{} & \ctrl{-5} & \ \ldots\ & \meter{}
      \end{quantikz}
    \end{adjustbox}
  }

  \caption{Comparison of different VQC architectures for our multi-agent experiments.}
  \label{fig:All_VQCs}
\end{figure*}
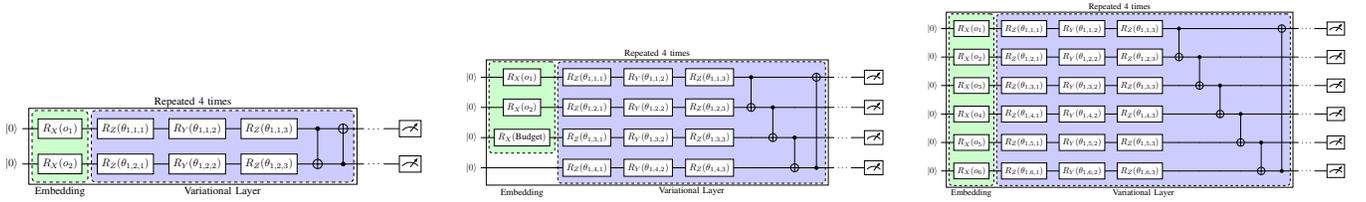

In our approach, each Q-Learning agent replaces the classical function approximator with a VQC. The observation is embedded into qubits and processed through $L$ layers of parameterized single-qubit rotations and circular CNOT entanglement, followed by measurement in the $Z$ basis. Measurements in $[-1,1]$ are rescaled to $[0,1]$ and multiplied by a trainable scaling factor to align with Q-values \cite{Skolik2022quantumagentsingym}.

We apply different embeddings depending on input dimensionality. \textit{Basis embedding} flips qubits based on binary features. \textit{Angle embedding} maps scalar values to rotation angles (e.g., $R_X(\cdot)$). \textit{Amplitude embedding} normalizes real-valued vectors and loads them into $M$ qubits via Mottonen state preparation \cite{mottonen2004transformationquantumstatesusing}. Optionally, we use data re-uploading \cite{PerezSalinas2020datareuploading}, re-embedding classical inputs in each variational layer to enhance expressivity.

\subsection{Matrix Game Architectures}
Matrix-game experiments (\cref{sec:matrix-games}) use a 4-layer VQC with Basis embedding, parameterized rotations, CNOT entanglement, and Pauli-Z measurement (\cref{fig:Standard VQC}--\cref{fig:RIAL VQC}). Outputs are rescaled to $[0,1]$ and multiplied by learnable scaling parameters with separate learning rate $\alpha_w$. The number of qubits depends on embedding and action-space size: baseline, MATE, AutoMATE, and MEDIATE variants use two qubits; Gifting variants use four (adding a binary gifting decision), and RIAL employs six qubits due to additional message bits. In Gifting\textsubscript{Budget}, one qubit encodes the agent’s budget (0–10 scaled to $[0,\pi]$).

Agents observe the joint previous actions $(a_{t-1,i}, a_{t-1,j})$; RIAL also includes message bits. Gifting actions remain hidden. In the first round, observations are random. The learning parameters are: $\epsilon$ decaying linearly from 0.3, $\alpha=0.001$, $\alpha_w=0.1$, $\gamma=0.9$. Each agent stores 50 transitions and trains after every episode using five mini-batches (size 5). Experiments run 2000 episodes of 50 steps. MATE tokens $x^{\text{MATE}}$ and $y^{\text{MATE}}$ equal 1; MEDIATE updates every 5 episodes ($\alpha_{\text{MEDIATE}}=0.1$). Gifts have $x^{\text{Gift}}=1$, and Gifting\textsubscript{Budget} assigns a budget of 10. Each setting is repeated 15 times (seeds 0–14).

\subsection{Harvest Game Architecture}
The Harvest Game represents a larger-scale Social Dilemma. A flattened $5\times5$ observation (100 features) is amplitude-embedded into 7 qubits. After $L$ variational layers, all qubits are measured and passed to a classical dense layer (size 6) corresponding to six possible actions. Both quantum and classical parameters share the same learning rate. Agents store 2500 transitions, train after each episode with batch size 100, use $\gamma=0.99$, and decay $\epsilon$ exponentially from 1 to 0.02: \(\epsilon_{\text{Current}} = \min\bigl(0.02,\;\epsilon_{\text{Initial}} \cdot 0.95^{\frac{100 \cdot \text{Current Episode}}{\text{Maximum Episodes}}}\bigr)\).
We test $L \in \{1,2,3,4\}$ and learning rates $\{0.01, 0.001, 0.0001\}$, and compare with a classical two-layer neural network (128 neurons per layer, ReLU). Each experiment spans 1000 episodes of 250 steps, repeated three times (seeds 0–2). These experiments assess quantum circuit scalability and efficiency in cooperative, visually rich environments.

\section{Experimental Setup}
\label{sec:experimental_setup}

\subsection{Matrix Games} \label{sec:matrix-games}
We evaluate the communication approaches described in \cref{sec:communication_protocols} and \cref{sec:vqc_architecture} in three classic two-player matrix games: the Prisoner’s Dilemma, the Stag Hunt, and the Game of Chicken. Each game is iterated over multiple rounds, and agents accumulate rewards across these rounds. All three games exemplify social dilemmas, highlighting conflicts between individual rationality and collective welfare.

In the Prisoner’s Dilemma, each player chooses to \emph{Cooperate} or \emph{Defect}, with payoffs satisfying \(T > R > P > S\) and \(2R > S + T > 2P\). Defection is the dominant strategy, yet mutual cooperation yields higher joint rewards. The Stag Hunt captures the trade-off between safety and cooperation: mutual hunting yields high payoffs, while unilateral foraging ensures modest but reliable returns. Finally, the Game of Chicken models brinkmanship, where each agent decides between cautious (\emph{chicken}) or risky (\emph{dare}) behavior, with payoffs following \(T > R > S > P\). No pure strategy dominates, forcing agents to balance cooperation and risk. The explicit payoff values follow standard formulations from the literature.

\subsection{Harvest Game}
\label{sec:harvest}
To evaluate scalability, we additionally test in the grid-based \emph{Harvest Game} \cite{NIPS2017_2b0f658c}, representing a spatial social dilemma. Agents navigate a grid to collect apples that regrow based on local density, requiring sustainable harvesting to avoid resource depletion. Each agent observes a $5 \times 5$ field centered on itself, encoding apples, other agents, and orientation. Agents can move, rotate, or stay still; tagging and firing actions remain unused.

We embed observations using seven qubits with \emph{amplitude embedding} and vary the number of variational layers ($L \in \{1,2,3,4\}$) and learning rates ($\alpha \in \{0.0001, 0.001, 0.01\}$). The quantum circuit outputs are post-processed by a small classical layer (six neurons) to estimate Q-values. As a baseline, we compare against a classical network with two hidden layers of 64 neurons each. Training proceeds for 1000 episodes of 250 steps, using an experience buffer of 2500 samples, batch size 100, $\epsilon$-greedy exploration decaying from 1.0 to 0.02, and discount factor $\gamma = 0.99$.

\subsection{Evaluation Metrics}
Emergent cooperation is assessed via (i) collective reward $C = \sum_{i,t} r_{t,i}$, (ii) frequency of mutual cooperation $FC = \sum_{t}\mathds{1}[a_{t,i}, a_{t,j}\text{ cooperative}]$, (iii) inequality $I = \sum_{t}|r_{t,i}-r_{t,j}|$, and (iv) auxiliary measures such as average token values or gifting frequency $FG = \frac{1}{|D|}\sum_{i,t}\mathds{1}[a_{t,i}\text{ is gift}]$. For final policy evaluation, we query trained models using greedy action selection ($\epsilon=0$) across all possible observations and report the fraction of cooperative actions.

\subsection{Training and Hyperparameters}
Quantum circuits are implemented in PennyLane using the lightning.qubit simulator \cite{asadi2024}. We compute exact Pauli-Z expectations and optimize parameters via the Adam optimizer using Mean Squared Error loss and adjoint differentiation. Parameters are initialized uniformly in $[-\pi, \pi)$.
For matrix games, agents observe both players’ previous actions, with $\epsilon$ initialized to 0.3 and linearly decayed over 2000 episodes of 50 steps. The learning rate is $\alpha=0.001$, output-scaling rate $\alpha_w=0.1$, discount $\gamma=0.9$, and buffer size 50 with batch size 5. Communication parameters include fixed MATE and gifting token values of 1.0, a MEDIATE update rate of 0.1, and two-bit RIAL messages. Each configuration is run for 15 random seeds.
For the Harvest Game, $\epsilon$ decays exponentially from 1.0 to 0.02 over 1000 episodes, each with 250 steps, using a buffer of 2500 and batch size 100. We test learning rates $\{0.0001, 0.001, 0.01\}$ and 1–4 variational layers, with $\gamma = 0.99$. All experiments are repeated for three seeds.

\section{Results}
\label{sec:results}


\subsection{Matrix Games}

\subsubsection{Iterated Prisoner's Dilemma}




\begin{figure*}[t]
  \centering
  \subfloat[Collective reward.\label{fig:ipd_reward}]{
    \includegraphics[width=0.31\linewidth]{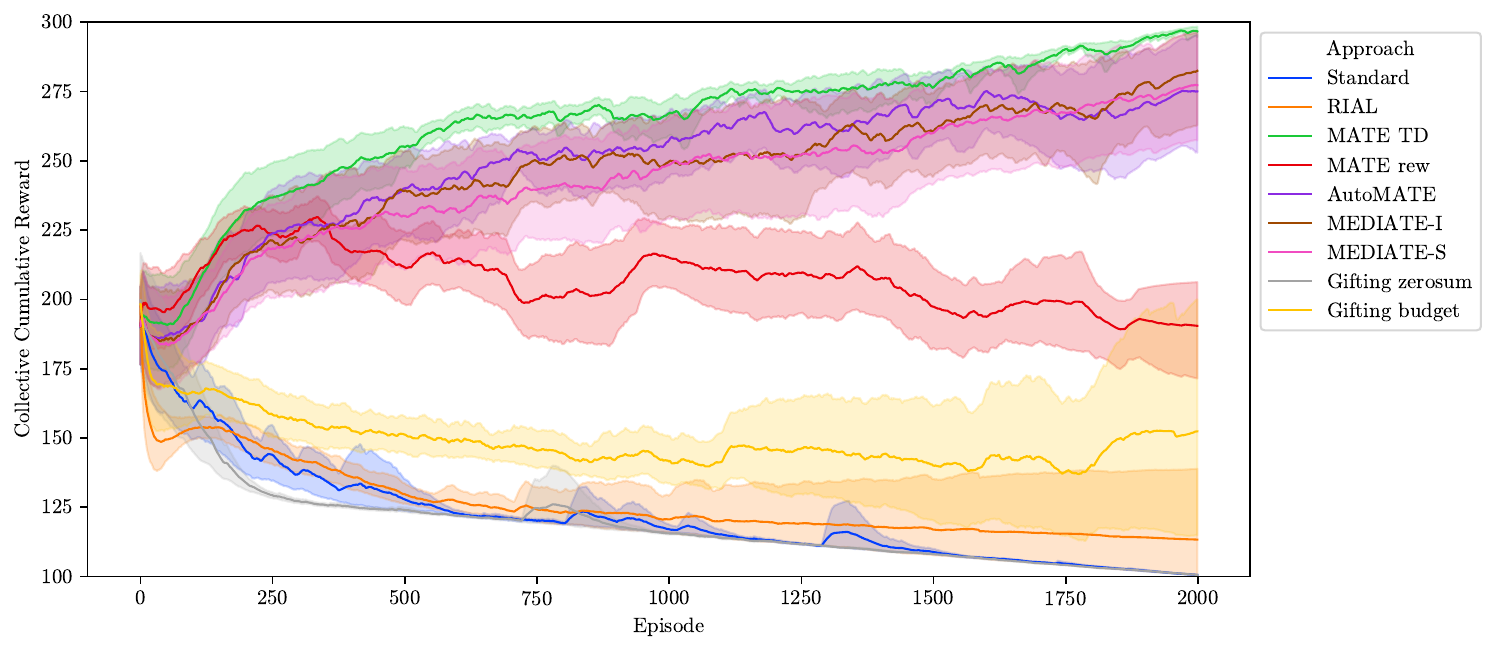}
  }
  \hfill
  \subfloat[Mutual cooperation frequency.\label{fig:Frequency of Mutual Cooperate Actions}]{
    \includegraphics[width=0.31\linewidth]{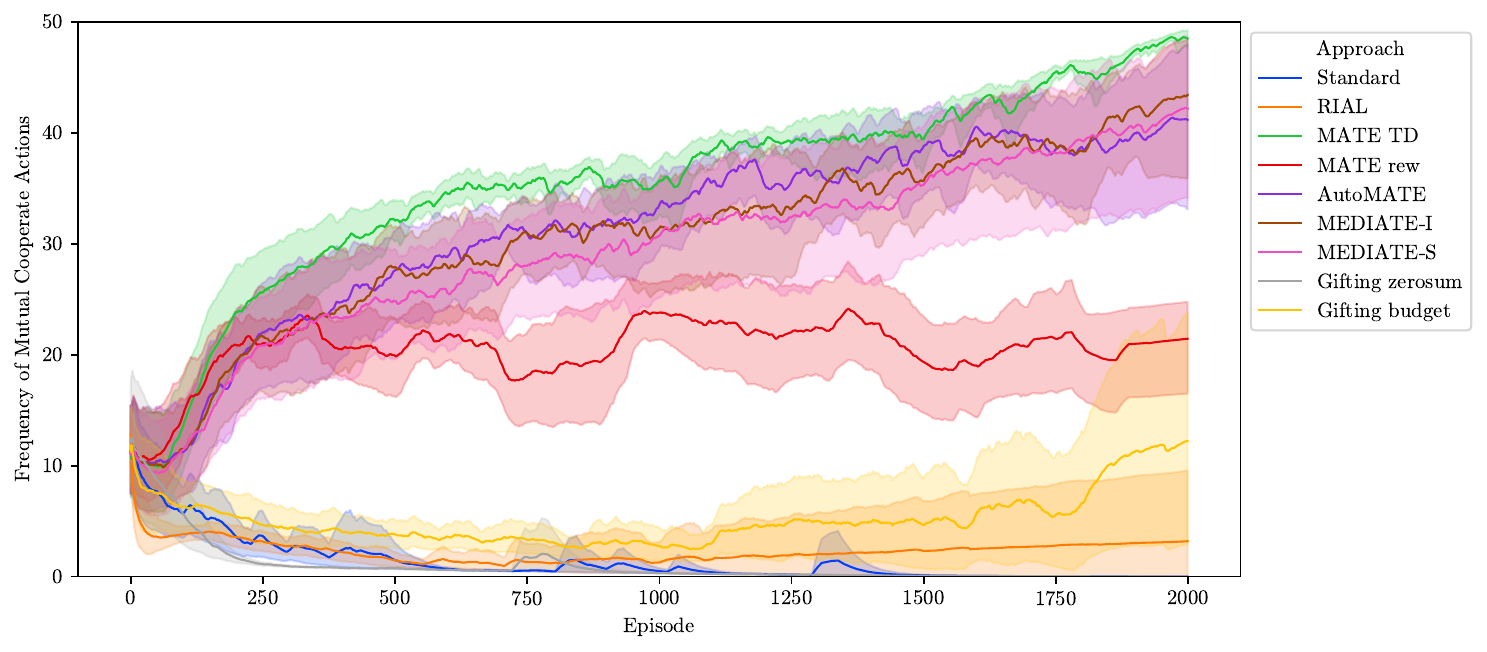}
  }
  \hfill
  \subfloat[Inequality.\label{fig:IPD Inequality}]{
    \includegraphics[width=0.31\linewidth]{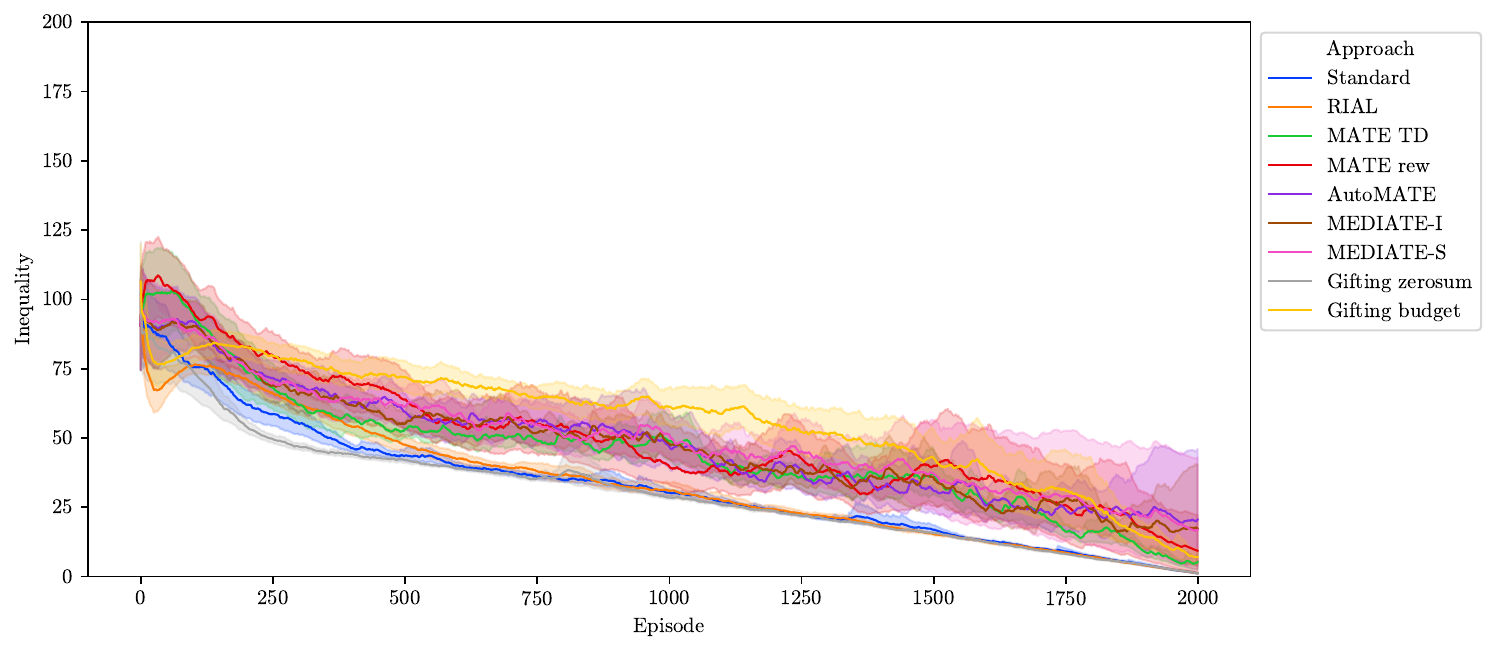}
  }

  \caption{Comparison of performance metrics in the Iterated Prisoner’s Dilemma (IPD) across all approaches: (a) collective cumulative reward, (b) cooperation frequency, and (c) inequality between agents.}
  \label{fig:IPD_combined}
\end{figure*}

We begin by examining the collective cumulative reward ($C$) across different approaches in the Iterated Prisoner's Dilemma. \cref{fig:ipd_reward} shows that MATE\textsubscript{TD} consistently achieved the highest $C$, reaching near-optimal performance early and maintaining low variance throughout training.
AutoMATE, MEDIATE-I, and MEDIATE-S follow closely, while MATE\textsubscript{rew} plateaus at a moderate level. Gifting\textsubscript{Budget} shows brief cooperation phases, but overall performs poorly. The baseline and Gifting\textsubscript{Zerosum} converge to mutual defection. RIAL marginally outperforms these but fails to achieve sustained cooperation.
The frequency of mutual cooperation ($FC$) in \cref{fig:Frequency of Mutual Cooperate Actions} supports this ranking. MATE\textsubscript{TD} leads with over 87\% mutual cooperation, followed by MEDIATE variants (73--83\%), while MATE\textsubscript{rew} is more conditional. Baseline, RIAL, and Gifting\textsubscript{Zerosum} mostly defect.
Inequality ($I$) is visualized in \cref{fig:IPD Inequality}. Gifting approaches initially show high $I$, especially Gifting\textsubscript{Budget}, indicating exploitative phases. MATE and MEDIATE approaches exhibit decreasing inequality, with MATE\textsubscript{TD} achieving the lowest.
%
MEDIATE approaches converge to derived token values of around 0.7, while MATE variants operate with fixed values of 1.0. The similarity suggests stable yet slightly less aggressive token incentives in MEDIATE.
Gifting frequencies ($FG$) reveal that Gifting\textsubscript{Zerosum} quickly abandons gifting, while Gifting\textsubscript{Budget} exhibits fluctuating but higher frequencies, possibly reflecting learning phases of gift–behavior association.
The final agent strategies confirm these trends: MATE\textsubscript{TD} agents almost always cooperate, MEDIATE variants cooperate frequently, and MATE\textsubscript{rew} behaves conditionally. Baseline, RIAL, and Gifting\textsubscript{Zerosum} rarely cooperate.




\subsubsection{Iterated Stag Hunt}

\begin{figure*}[t]
  \centering
  \subfloat[Collective reward\label{fig:Stag Hunt Low Risk Reward}]{
    \includegraphics[width=0.31\linewidth]{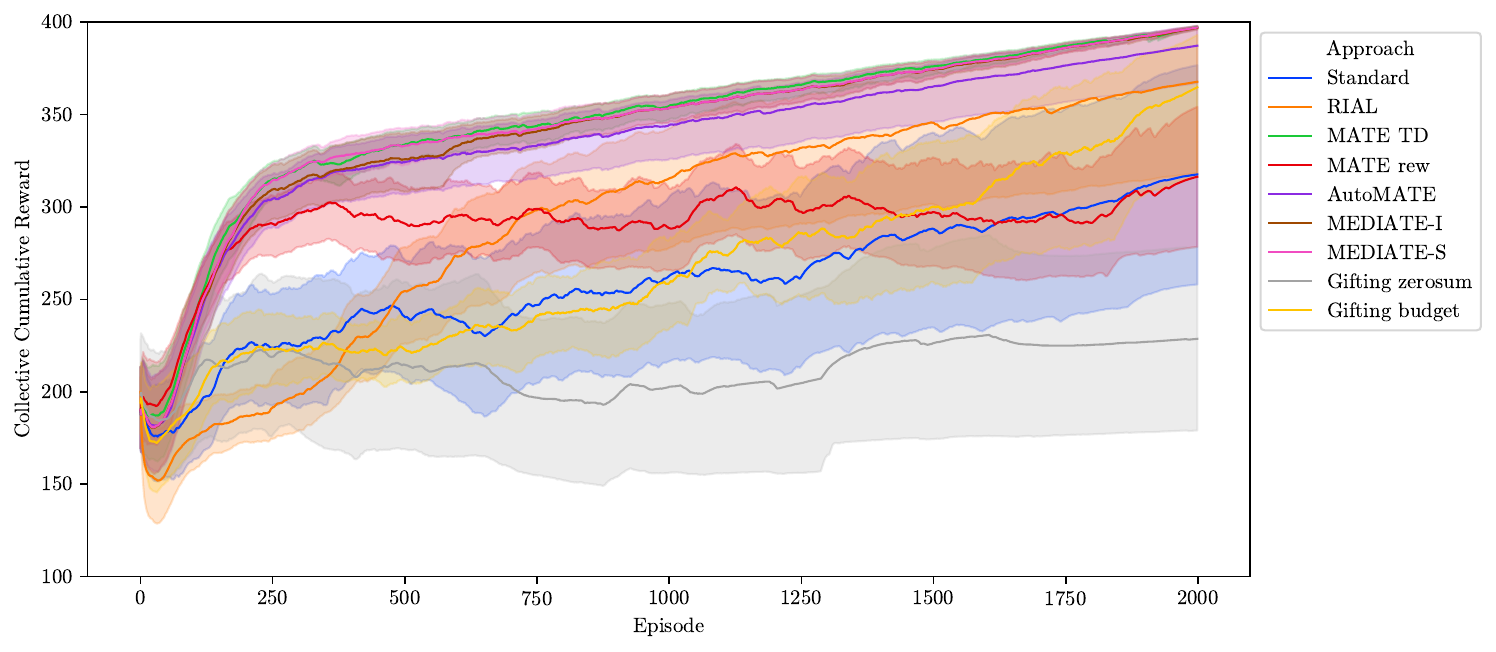}
  }
  \hfill
  \subfloat[Mutual hunt frequency\label{fig:Stag Hunt Low Risk Frequency of Mutual Hunt Actions}]{
    \includegraphics[width=0.31\linewidth]{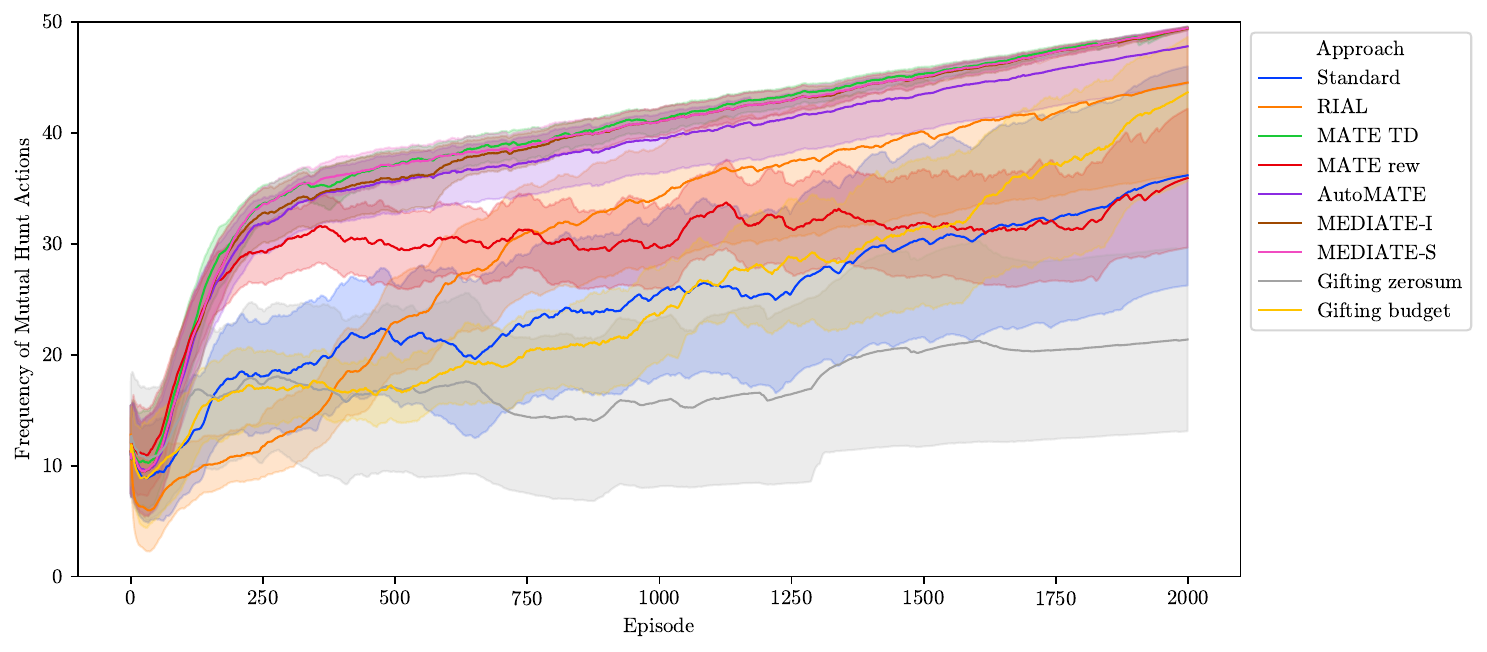}
  }
  \hfill
  \subfloat[Inequality\label{fig:Stag Hunt Low Risk Inequality}]{
    \includegraphics[width=0.31\linewidth]{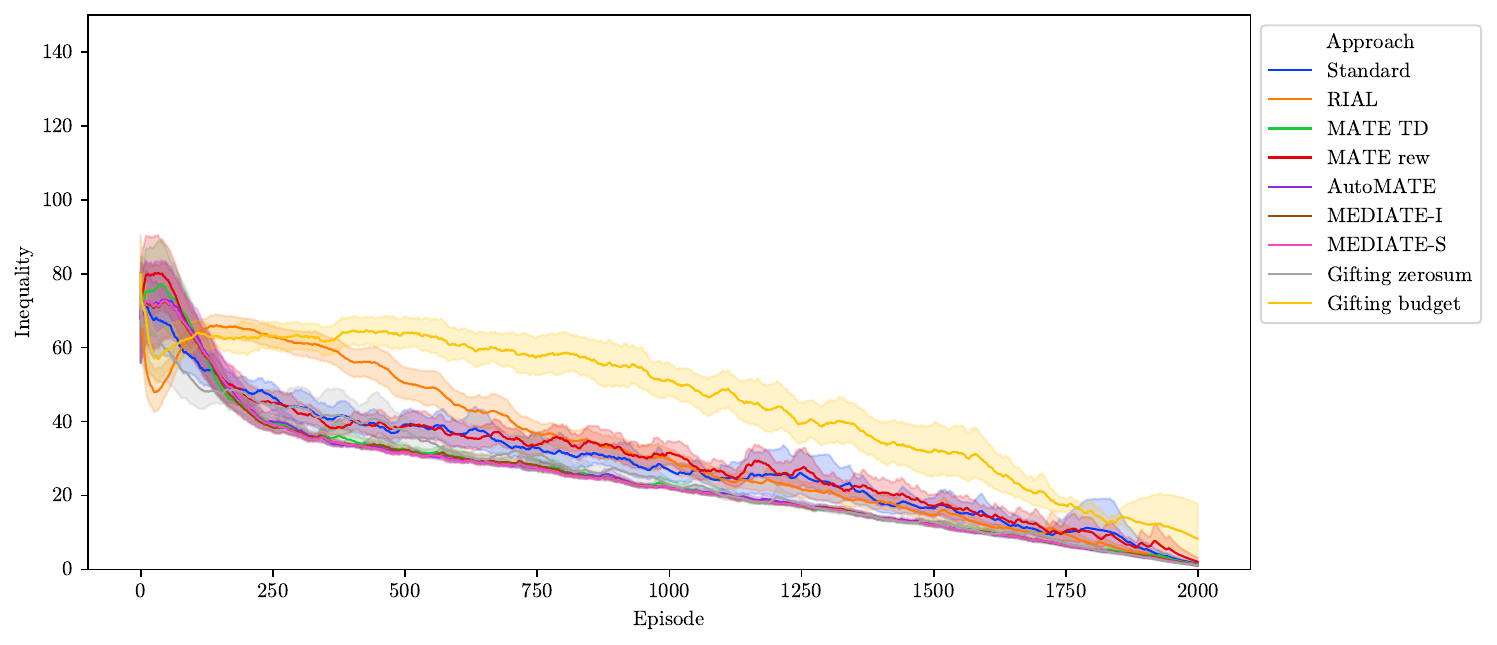}
  }

  \caption{Performance metrics in the Iterated Stag Hunt (low-risk variant): (a) collective cumulative reward, (b) mutual hunt frequency, and (c) inequality between agents.}
  \label{fig:StagHunt_combined}
\end{figure*}

In the Iterated Stag Hunt, collective rewards (\cref{fig:Stag Hunt Low Risk Reward}) and mutual hunt frequency (\cref{fig:Stag Hunt Low Risk Frequency of Mutual Hunt Actions}) indicate near-optimal cooperation from MATE\textsubscript{TD}, MEDIATE-I, and MEDIATE-S. AutoMATE also performs well, but MATE\textsubscript{rew} stagnates mid-training.
Inequality in \cref{fig:Stag Hunt Low Risk Inequality} is lowest for the top-performing methods. Gifting\textsubscript{Budget} shows more exploitation phases. Token values again settle around 0.7. Our experiments show that gifting is eventually abandoned, especially in Gifting\textsubscript{Zerosum}.
Learned strategies show most approaches learned to hunt after observing cooperation. Gifting\textsubscript{Zerosum} failed to develop this behavior, while RIAL and Gifting\textsubscript{Budget} reached partial success.

\subsubsection{Iterated Game of Chicken}

\begin{figure*}[t]
  \centering
  \subfloat[Collective reward\label{fig:Chicken Reward}]{
    \includegraphics[width=0.31\linewidth]{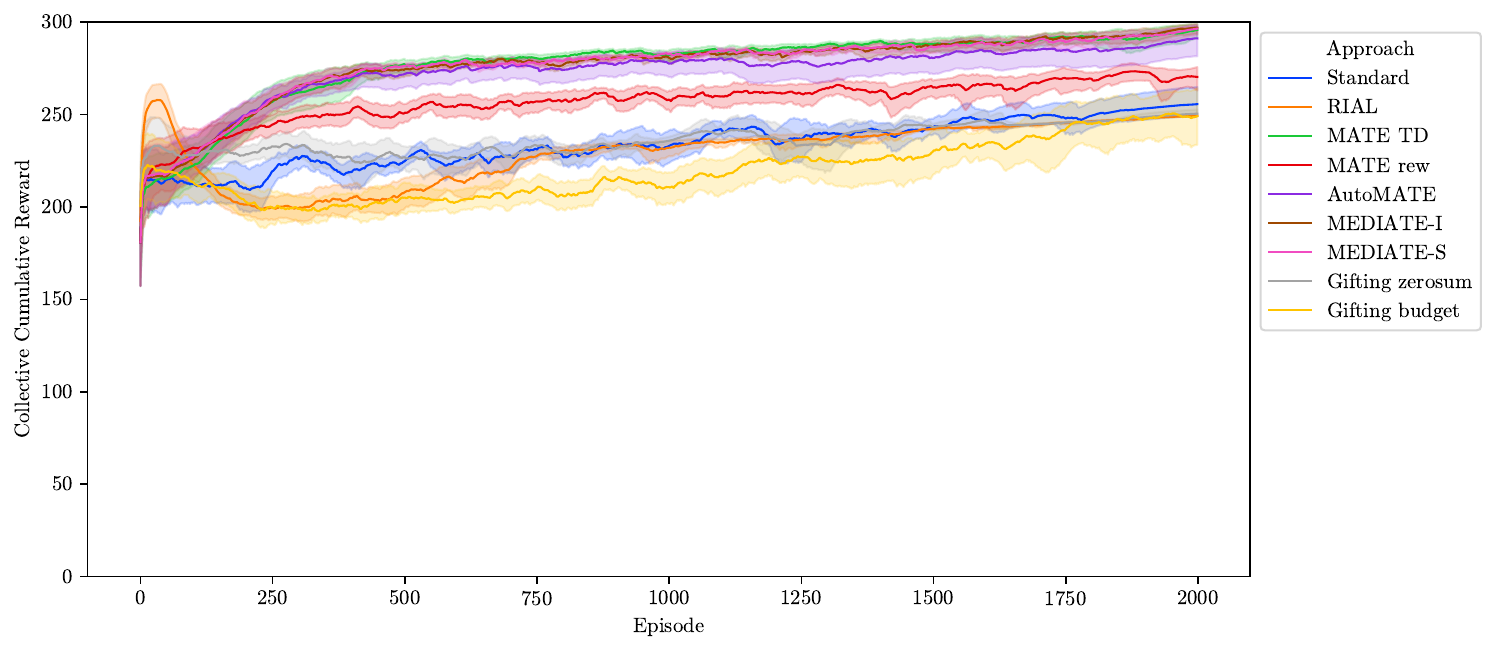}
  }
  \hfill
  \subfloat[Mutual chicken frequency\label{fig:Frequency of Mutual Chicken Actions}]{
    \includegraphics[width=0.31\linewidth]{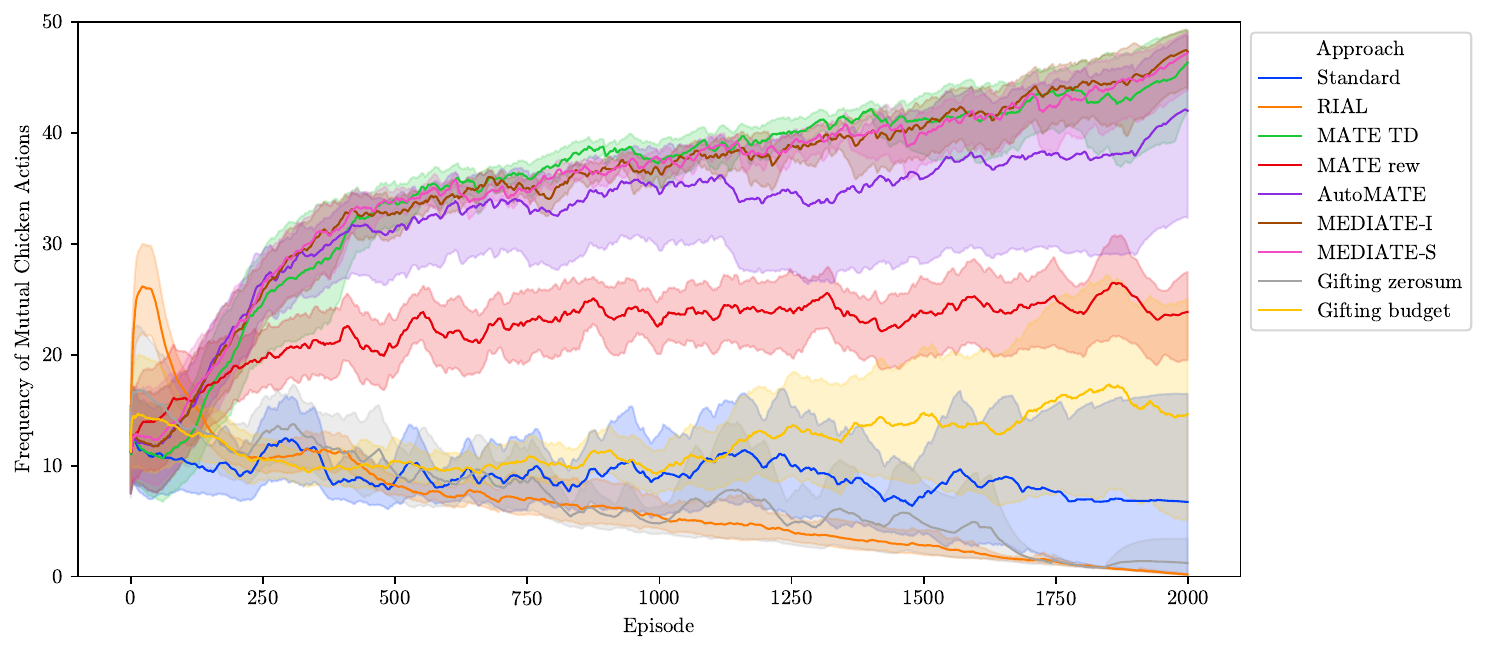}
  }
  \hfill
  \subfloat[Inequality\label{fig:Chicken Inequality}]{
    \includegraphics[width=0.31\linewidth]{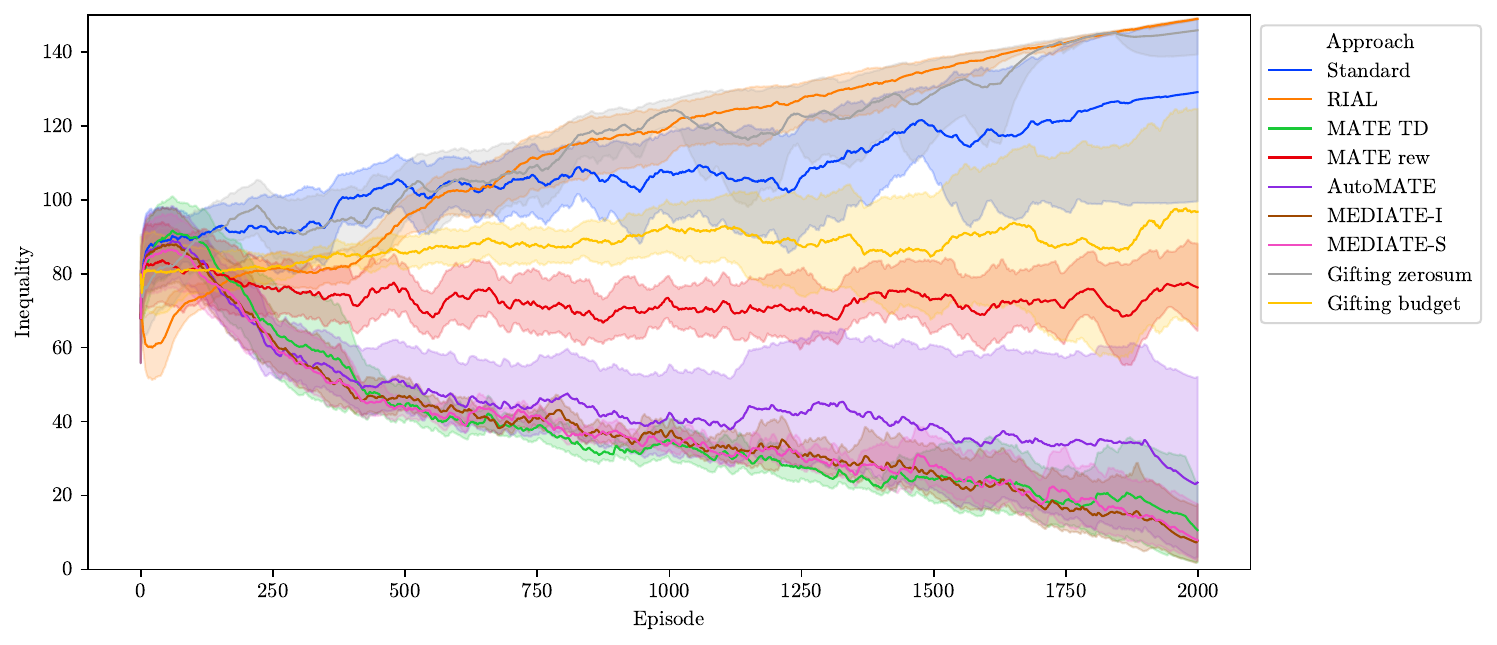}
  }

  \caption{Performance metrics in the Iterated Game of Chicken: (a) collective cumulative reward, (b) frequency of mutual chicken actions, and (c) inequality between agents.}
  \label{fig:Chicken_combined}
\end{figure*}

\cref{fig:Chicken Reward} shows that MATE\textsubscript{TD}, MEDIATE-I, and MEDIATE-S again achieve high collective reward. AutoMATE closely follows. MATE\textsubscript{rew} provides moderate success; others struggle. Inequality (\cref{fig:Chicken Inequality}) reflects exploitative equilibria in RIAL and Gifting\textsubscript{Zerosum}, both reaching near-maximal $I$.
Mutual chicken actions (\cref{fig:Frequency of Mutual Chicken Actions}) increase steadily in top approaches. Gifting\textsubscript{Budget} improves late; Gifting\textsubscript{Zerosum} and RIAL stagnate.
Token values match previous games. Gifting frequency again drops sharply.
The final strategies show strong cooperation in MATE\textsubscript{TD}, MEDIATE, and AutoMATE. MATE\textsubscript{rew} shows moderate levels. RIAL and Gifting variants exhibit low cooperation.

\subsection{Harvest Game Results}

\begin{figure}[t]
  \centering
  \includegraphics[width=\linewidth]{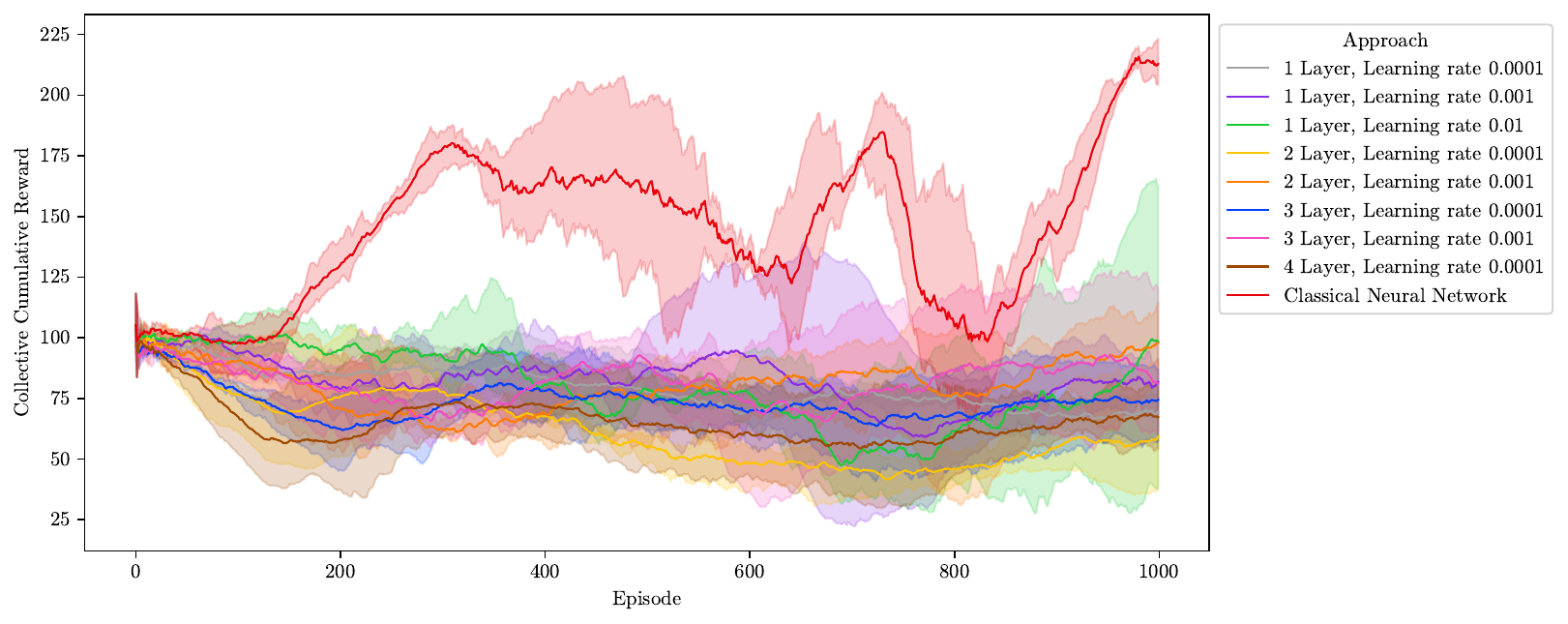}
  \caption{Collective cumulative reward in the Harvest Game.}
  \label{fig:harvest_reward}
\end{figure}

The performance results in the Harvest Game (\cref{fig:harvest_reward}) show that all VQC variants struggled to match the classical neural network baseline. The classical network displayed characteristic learning patterns with performance valleys around episodes 300 and 600, likely representing phases of over-harvesting followed by adaptation. These valleys are typical of the tragedy of the commons scenario, where agents initially deplete resources before learning sustainable harvesting strategies.
Among the VQC variants, the 1-layer circuit with learning rate 0.01 maintained the highest performance through most of training. The 2-layer circuit with learning rate 0.001 showed promising improvement in the middle phases but couldn't sustain this advantage. The 1-layer circuit with learning rate 0.001 exhibited more oscillatory behavior, with periods of higher performance followed by decline.
Deeper circuits (3-4 layers) generally performed worse, suggesting that additional parametrized layers did not improve representation capacity for this task and may have introduced optimization challenges. The lowest performing variant was the 2-layer circuit with learning rate 0.0001, which showed consistent performance degradation throughout training.
Notably, all VQC variants consistently performed below the random policy baseline (represented by the initial performance at episode 0) for significant portions of training. This suggests fundamental challenges in quantum representation learning for complex observation spaces like those in the Harvest Game. 

The poor performance of quantum agents in the Harvest Game can be attributed to several factors:
The high-dimensional observation space of the Harvest Game ($5 \times 5$ multi-channel grid) must be flattened and normalized for Amplitude embedding, potentially losing structural information. Despite using seven qubits, the VQC's representational capacity may be insufficient for learning complex Q-functions required for multi-step planning and sustainable resource management. Variational quantum circuits face challenges in optimization, including barren plateaus where gradients vanish exponentially with system size. This problem is particularly acute with deeper circuits and complex encodings. The Amplitude encoding may produce quantum states that are difficult to distinguish, limiting the VQC's ability to learn appropriate Q-values for different observations. These results highlight important limitations of current quantum learning approaches for complex cooperative tasks and suggest that further research is needed to design quantum architectures specifically optimized for multi-agent settings with high-dimensional observation spaces.

\section{Conclusion}
\label{sec:conclusion}

In this paper, we studied emergent cooperation in QMARL by adapting eight classical MARL communication approaches across three SSDs. Our results show that communication protocols can foster cooperation among quantum Q-Learning agents, confirming the feasibility of communication-based coordination in quantum settings.
Temporal-difference-based monotonic improvement methods—MATE\textsubscript{TD}, AutoMATE, MEDIATE-I, and MEDIATE-S—achieved the highest cooperation across all dilemmas, highlighting the importance of long-term reasoning in mitigating fear and greed incentives. Token-based protocols outperformed other approaches, with MATE\textsubscript{TD} showing the most consistent results. MEDIATE variants successfully derived effective token values (around 0.7) and benefited from their consensus mechanism, which notably improved cooperation in exploitative games like the Game of Chicken.
RIAL’s simple message passing proved context-dependent—ineffective in the Prisoner’s Dilemma but successful in the Stag Hunt—suggesting that unincentivized communication only aids coordination when partial interest alignment exists. Peer rewarding showed limited benefit: Gifting\textsubscript{Budget} modestly improved cooperation, while Gifting\textsubscript{Zerosum} often ceased gifting due to penalties, underscoring the sensitivity of reward-sharing mechanisms.
Overall, our findings demonstrate that classical MARL communication strategies can be effectively transferred to quantum agents, with token-based protocols offering the most robust cooperation. Future work should explore more complex environments, study the impact of different VQC architectures, and design communication protocols exploiting quantum properties such as entanglement, as well as test scalability to larger agent populations.

\bibliographystyle{ieeetr}
\bibliography{main}


\end{document}